\documentclass[letterpaper, 10 pt, conference]{ieeeconf}  

\IEEEoverridecommandlockouts                              
\usepackage[noadjust]{cite}
\usepackage{mathtools,amssymb}
\usepackage{siunitx}

\DeclareSIUnit\sq{\ensuremath{\Box}}

\usepackage{graphics} 
\usepackage{amsmath} 
\usepackage{amssymb}  

\title{\LARGE \bf
Optical Response of Lumped-Element Kinetic-Inductance Detector Arrays
}

\author{Shibo Shu, Martino Calvo, Johannes Goupy, Andrea Catalano, Aurelien Bideaud,\\Alessandro Monfardini, Samuel Leclercq and Eduard F.C. Driessen
\thanks{*This work has been partially funded by the LabEx FOCUS ANR-11-LABX-0013.}
\thanks{S. Shu, S. Leclercq and E.F.C. Driessen are with the Institut de RadioAstronomie Millim\'{e}trique, 300 rue de la Piscine, 38406 Saint Martin d$'$H\`{e}res, France.
        {\tt\small shu@iram.fr}}
\thanks{M. Calvo, J. Goupy, A. Bideaud and A. Monfardini are with Institut N\'{e}el, CNRS and Universit\'{e} Grenoble Alpes, 25 avenue des Martyrs, 38042 Grenoble, France.}
\thanks{A. Catalano is with the Laboratoire de Physique Subatomique et de Cosmologie, Universit\'{e} Grenoble Alpes and the Institut N\'{e}el, CNRS, 53 avenue des Martyrs, Grenoble, France.}
}

\begin{document}

\maketitle
\thispagestyle{empty}
\pagestyle{empty}

\begin{abstract}
We present an analysis of the optical response of lumped-element kinetic-inductance detector arrays, based on the NIKA2 1mm array. This array has a dual-polarization sensitive Hilbert inductor for directly absorbing incident photons. We present the optical response calculated from a transmission line model, simulated with HFSS and measured using a Fourier transform spectrometer. We have estimated the energy absorbed by individual component of a pixel, such as the inductor. The difference between the absorption efficiencies is expected to be 20\% from the simulations. The Fourier-transform spectroscopy measurement, performed on the actual NIKA2 arrays, validates our simulations. We discuss several possible ways to increase the absorption efficiency. This analysis can be used for optimization of the focal plane layout and can be extended to other kinetic inductance detector array designs in millimeter, sub-millimeter and terahertz frequency bands. 

\end{abstract}


\section{Introduction}
To detect weak astronomical signals, superconducting devices such as kinetic inductance detectors (KIDs)~\cite{Day:2003a} and transition edge detectors (TESs)~\cite{Irwin:2005a}, are widely developed for many astrophysics applications in millimeter, sub-millimeter, and far-infrared wavelength bands~\cite{Monfardini:2010a,Sayers:2014a,Adam:2018a,Ferrari:2016a,Hubmayr:2015a}. For total-power detection, usually a large number of pixels are required to increase the observation efficiency and mapping speed~\cite{Farrah:2017a}. For example, the SCUBA2 instrument~\cite{Holland:2013a} has \num{e4} pixels and the OST satellite~\cite{Fortney:2018a} requires \num{e5} pixels for its far-infrared imager. In the astronomical detection field, KIDs have attracted great interest and have been developed rapidly in the last decade. Compared with TESs, the main advantages of KIDs are their simple structure and the intrinsic frequency domain multiplexing property. KIDs are based on superconducting microresonators~\cite{Zmuidzinas:2012a}, coupled capacitively or inductively to a feedline for frequency multiplexing. When the energy of incident photon is larger than the superconducting gap ($h\nu>2\Delta$), the Cooper pairs are broken and quasi-particles are generated. This increases the kinetic inductance and the loss in the superconductor. The increased kinetic inductance shifts the resonance frequency and the increased loss decreases the resonance dip. Both can be read out from the resonance curve. Using a single feedline, hundreds or even thousands of KIDs can be read out simultaneously.

In the millimeter and terahertz fields, there are mainly two kinds of KIDs, the distributed and lumped-element ones~\cite{Zmuidzinas:2012a,Mazin:2009a,Baselmans:2012b}. Distributed KIDs have distributed capacitance and inductance in the form of a section of a planar transmission line with a resonant frequency dependent on the length. The length of the waveguide determines the resonance frequency. Usually a resonant on-chip antenna is designed at the end of the resonator. The lumped-element KID (LEKID)~\cite{Doyle:2008a} consists of a lumped-element inductor and capacitor, which separates the photon-sensitive part, the inductor, and the frequency-tuning part, the capacitor. This separation allows more flexible design for the inductor. The inductor can be located after a horn for coupling signal~\cite{McCarrick:2014a,Hubmayr:2013a}, utilizing the well-studied optical response of horn. The inductor can also be designed to absorb the free-space electromagnetic wave directly without any antenna~\cite{Doyle:2008a}. In this paper, the analysis is based on this bare LEKID design.

To improve the sensitivity of KIDs, it is necessary to evaluate and maximize the amount of energy absorbed by the inductor. For distributed KIDs, the responsivity depends on the current distribution along the planar transmission line~\cite{Mazin:2005a}. It can be improved by using a hybrid structure with a high-Tc superconductor to restrict the photons to be absorbed in the most-sensitive part~\cite{Janssen:2013a}. For horn-coupled LEKIDs, only the inductor part is irradiated and the optical coupling is simply determined by the impedance matching between the inductor and incident wave~\cite{Hubmayr:2013a}. Roesch et al.~\cite{Rosch:2013a} have investigated the optical response of a bare LEKIDs array in the \SI{2}{mm} atmospheric band. In their model, the array consists of a set of periodic pixels only, and the readout line is not considered. The absorption efficiency is calculated from the return loss measured using a vector network analyzer. In their analysis the simulation and measurement matched well, however, the connection between the measured response of KID and the absorption in the pixel is missing.

In this paper, we demonstrate a thorough electromagnetic simulation analysis of a bare LEKID array, such as used in the NIKA2 instrument~\cite{Adam:2018a}. The energy absorbed in the inductor and in other components are separated using the Surface Loss function in the Field Calculator of HFSS~\cite{HFSS}. The simulation results are compared to measurements using a Fourier transform spectrometry (FTS). We find that the LEKID array has a large difference of absorption efficiency for the two incident polarizations. Several configurations are simulated for understanding the absorption in the different components of the pixel. This analysis can be used for optimization of the design of the focal plane components of LEKID arrays. Extending the analysis to other KID designs is also possible.

\section{Array Layout} \label{sec:layout} 
\begin{figure}
\centering
\includegraphics[width=0.45\textwidth,trim={15.3cm 3cm 13cm 3cm},clip]{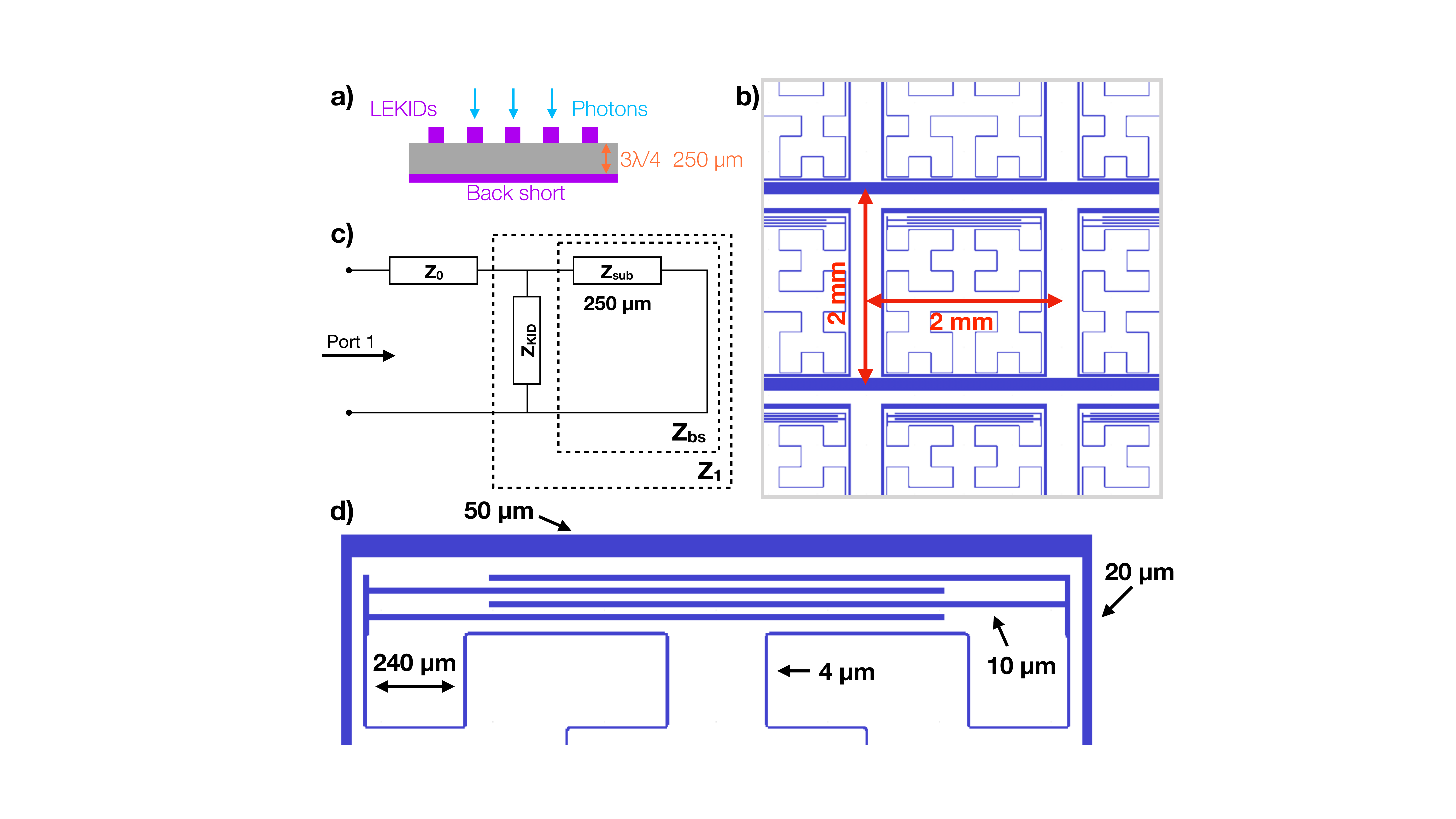}
\caption{a) Cut view of the LEKIDs array in the photon incident direction. b) Schematic drawing of the NIKA2 1mm array. The pitch size is \SI{2}{mm}. c) Transmission line model of the optical coupling in (a). The LEKID, mainly the inductor, is represented by a sheet impedance $Z_{\textrm{KID}}$. d) The design geometries of the LEKID pixel. The frame size is $\SI{1.6}{mm}\times \SI{1.7}{mm}$ and the readout line width is \SI{120}{\um}. }
\label{fig:TML}
\end{figure}
The NIKA2 1mm LEKIDs array~\cite{Adam:2018a} (Fig.~\ref{fig:TML}) is simulated and measured in this paper. This array consists of $1140$ pixels read out by $8$ microstrip feedlines. The size of the array is a circle with radius of \SI{40}{mm} corresponding to a \SI{6.5}{\arcmin} field-of-view on the IRAM 30-m telescope. The pixel pitch size and the inductor size are $2\times \SI{2}{mm^2}$ and $1.6\times \SI{1.5}{mm^2}$, respectively. Each pixel has the same inductor design with different capacitor finger lengths for different designed resonance frequencies. The back of the wafer is covered with \SI{200}{nm} thick aluminium, and acts at the same time as the ground plane for the MS readout feedline and as the backshort to optimize the optical coupling.

This array uses the bare LEKID design, which has no antenna structure on the focal plane. The incident light illuminates the array directly. The detection band of the LEKIDs is determined by the backshort distance, which is the thickness of the silicon substrate. The band central frequency (wavelength) is designed to be \SI{260}{\GHz} (\SI{1.15}{\mm}), the central frequency of the \SI{1}{mm} atmospheric window. The backshort is designed to be $3\lambda_{\textrm{Si}}/4=\SI{250}{\um}$, where the wavelength in silicon $\lambda_{\textrm{Si}}=\lambda/\sqrt{\epsilon_r}=\SI{333}{\um}$, the wavelength in free space $\lambda=\SI{1.15}{\mm}$ and the relative permittivity of silicon $\epsilon_r=11.9$. The bandwidth is expected from \SIrange{230}{290}{\GHz}, limited by the $3\lambda/4$ backshort distance. Using a $\lambda/4$ backshort will increase the bandwidth to fill the entire \SI{1}{mm} atmospheric window.

The inductor is designed with a 3rd-order Hilbert curve~\cite{Roesch:2012a} for dual-polarization sensitivity. The inductor width and the line spacing are designed as $s=\SI{4}{\um}$ and $w=\SI{240}{\um}$, respectively, shown in Fig.~\ref{fig:TML}(d). Since the wavelength is much larger than the structure, the inductor can be roughly modeled as a sheet impedance $Z_{\textrm{KID}}$. The optical coupling can be modeled using the transmission line model~\cite{Roesch:2014a}, shown in Fig.~\ref{fig:TML}(c). The impedance of the backshort can be expressed as
\begin{equation}
Z_{\textrm{bs}}=jZ_{\textrm{sub}}\tan (\beta l),
\end{equation}
where $\beta=2\pi/\lambda_{\textrm{Si}}$ and $l$ is the backshort distance (\SI{250}{\um}). The effective impedance of the KIDs together with the backshort is
\begin{equation}
Z_{\textrm{1}}=\frac{1}{\frac{1}{Z_{\textrm{KID}}}+\frac{1}{Z_{\textrm{bs}}}}.
\end{equation}
Assuming the reactance is zero, the impedance of KID is
\begin{equation}
Z_{\textrm{KID}}=R_{\sq}/(s/w).
\end{equation}
Then the absorption efficiency is calculated as
\begin{equation}
\label{eqn:model}
\textrm{absorption efficiency} = 1-\left | S11 \right |^2=1-\left | \frac{Z_{\textrm{0}}-Z_{\textrm{1}}}{Z_{\textrm{0}}+Z_{\textrm{1}}}\right |^2,
\end{equation}
where the vacuum impedance $Z_0=\SI{377}{\ohm}$. Given the sheet resistance $R_{\sq}=\SI{1.6}{\ohm\per\sq}$ of the used \SI{20}{\nm} aluminium film, $Z_{\textrm{KID}}=\SI{96}{\ohm}$ and the maximum absorption is 64.7\% at the band center. This is only a simple estimation for the absorption efficiency. Actually, away from the band center, the absorption efficiency largely depends on the reactance of $Z_{\textrm{KID}}$, which is usually not zero and difficult to be evaluated analytically. Therefore, we use electromagnetic simulation as a useful tool for focal plane array design.

\section{Simulation Method} \label{sec:simulation_method}

For a bare LEKID array, the pixels are arranged repeatedly on the focal plane and the differences of capacitor finger lengths are relatively small compared to the wavelength. Therefore, the full array can effectively be approximated as a single pixel with periodic boundary conditions. This method dramatically decreases the simulation time and enables the analysis of the optical response of the array. All simulations presented in this paper are simulated in HFSS~\cite{HFSS}.

\begin{figure*}[ht]
\begin{center}
\begin{tabular}{c}
\includegraphics[width=0.9\textwidth,trim={0cm 8.5cm 0cm 8.5cm},clip]{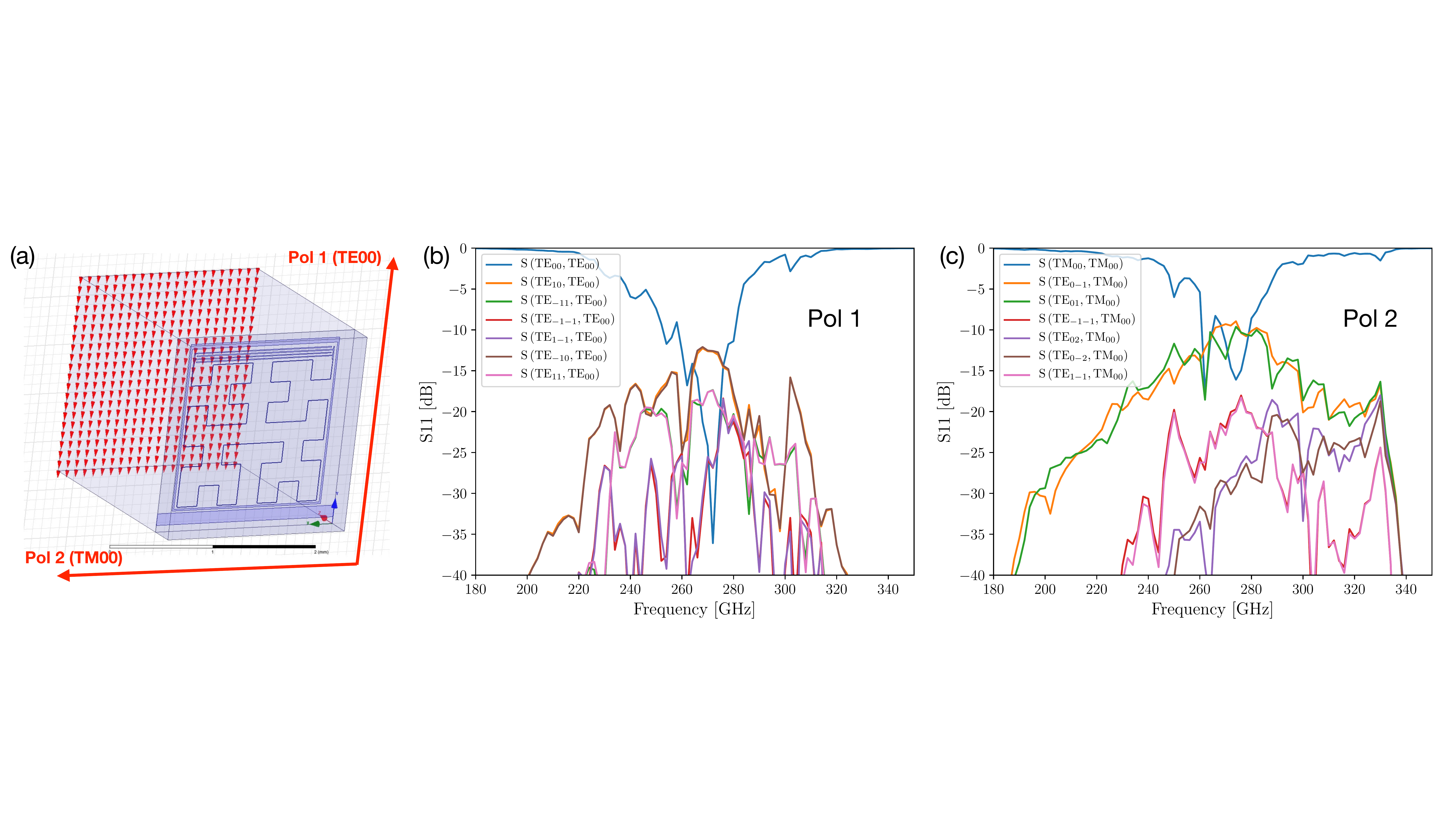}
\end{tabular}
\end{center}
\caption{a) The simulation model of a single pixel in HFSS. Two polarizations ($\textrm{TE}_{00}$ and $\textrm{TM}_{00}$) are stimulated as the incident signal. The arrows show the direction of the electric field in the two dominant modes. b) S-parameter results for a $\textrm{TE}_{00}$ excitation (Pol 1). Only the modes with a maximum return loss higher than \SI{-20}{dB} are plotted for clarity (same for (c)). The $\textrm{TE}_{10}$ and $\textrm{TE}_{-10}$ modes have more reflect energy than other higher modes. c) S-parameter results for a $\textrm{TM}_{00}$ excitation (Pol 2). $\textrm{TE}_{0-1}$ and $\textrm{TE}_{01}$ modes reflect more energy than other higher modes.}
\label{fig:Spara}
\end{figure*}

The pixel is built with a \SI{20}{nm} thick structure, which is the smallest size in our model. All the metal structures are assigned as either impedance boundaries of \SI{1.6}{\ohm\per\sq} or perfect electric conductor (PEC), depending on the configurations we are considering. The \SI{200}{nm} Al backshort is simplified as a PEC boundary assigned to the backside of the substrate, since the absorption efficiency is negligible due to the impedance mismatch. The only excitation port is Port 1, located on top of the model. The four sides of the model are assigned as periodic boundaries (Fig.~\ref{fig:Spara}(a)).

In our instrument, the incident direction of the focused radiation is perpendicular to the focal plane for all positions, due to our telecentric lens system design~\cite{Adam:2018a}. This is similar to a plane wave illumination. In the model, a Floquet port is assigned to Port 1 for exciting plane waves. Considering the simulating frequency from \SIrange{150}{350}{\GHz}, 18 modes are included in the analysis. Two TEM modes, $\textrm{TE}_{00}$ and $\textrm{TM}_{00}$ with $y$ and $x$-axis polarizations, respectively, are simulated as the incident signal.

The absorption efficiency in the pixel is calculated as
\begin{equation}
\label{eqn:S11}
\textrm{AE}_{\textrm{S11}} = 1 - \sum_{\textrm{All modes}} \left |\textrm{S11}\right |^2,
\end{equation}
where S11 are the return losses.

\subsection{Absorption efficiency calculation using Surface Loss}
\label{sec:surface_loss}
The absorption efficiency calculated using Eqn.~\ref{eqn:S11} includes the energy absorbed by all components. However, the most sensitive part of a LEKID is the inductor, where the created quasi-particles give rise to the change of readout signal. To extract the energy absorbed by this individual component, we use the surface loss density function in the Field Calculator. The surface loss density is calculated as
\begin{equation}
p_s=Re(\vec{P} \cdot \vec{n}),
\end{equation}
where $\vec{P}$ is the Poynting vector and $\vec{n}$ is the vector normal to the surface. The surface loss is then calculated by integration of the surface loss density over all surfaces of a component,
\begin{equation}
\textrm{surface loss}=\int_{S} p_s dS.
\end{equation}
The absorption efficiency calculated from surface loss, $\textrm{AE}_{\textrm{SL}}$, is given by
\begin{equation}
\label{eqn:sl}
\textrm{AE}_{\textrm{SL}}=\frac{\textrm{surface loss}}{\textrm{incident power}},
\end{equation}
where the incident power is known. In the simulation, the pixel is geometrically separated into four components: the inductor, capacitor, frame and readout line. Each component has a defined thickness  (\SI{20}{nm}). This analysis allows to evaluate the absorption efficiency in each component. The difference between the absorption efficiency calculated from $\textrm{S11}$ and surface loss is negligible and discussed in Sec.~\ref{sec:results 1}

\section{Simulation and Measurement of LEKID Array} \label{sec:results}

\subsection{Absorption of NIKA2 1mm array}
\label{sec:results 1}
\begin{figure*}[ht]
\begin{center}
\begin{tabular}{c}
\includegraphics[width=0.9\textwidth,trim={0cm 8.5cm 0cm 8.5cm},clip]{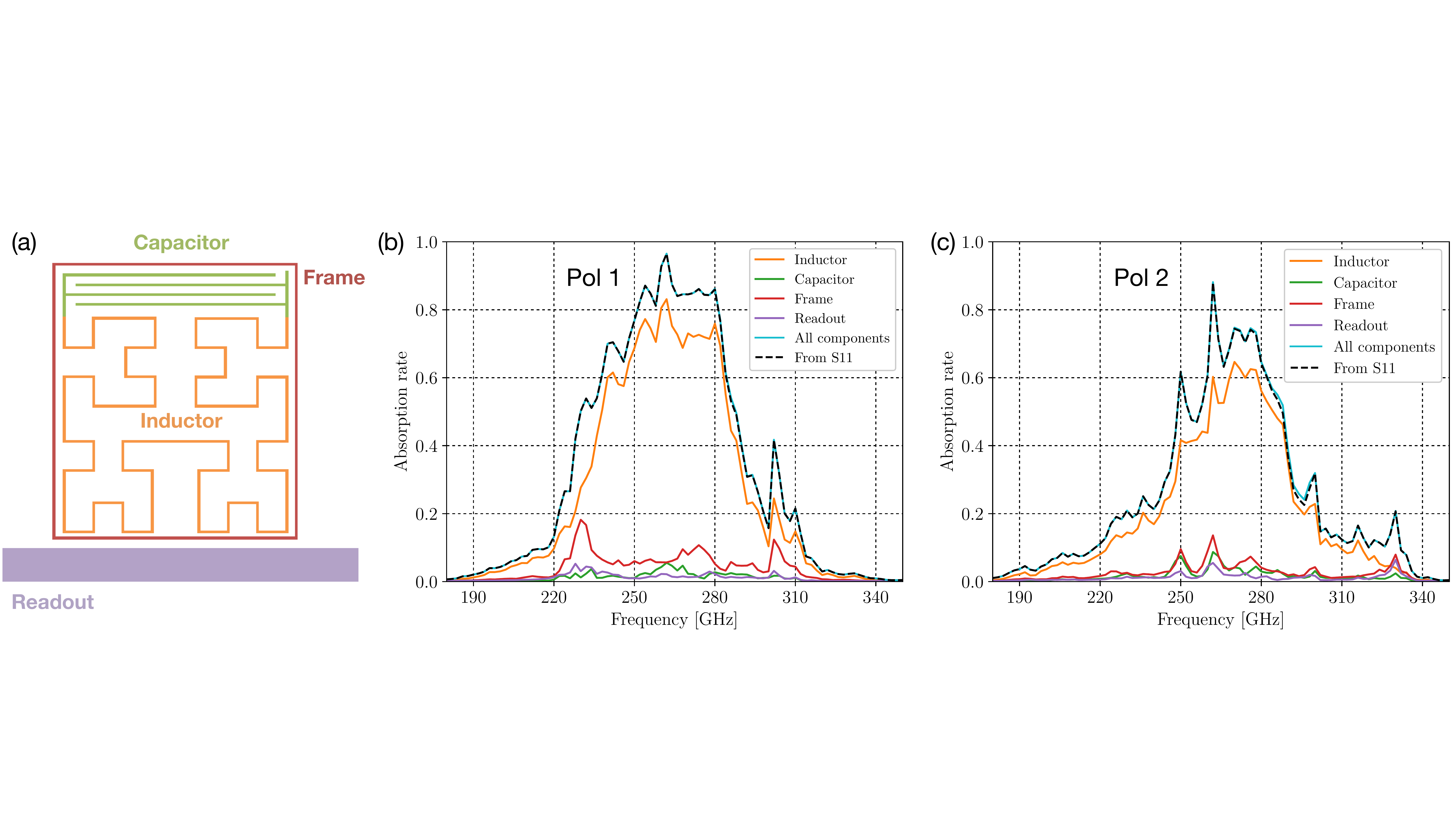}
\end{tabular}
\end{center}
\caption{a) A single pixel in simulation is separated into four components, the inductor capacitor, frame and readout line. The absorption efficiency of each component is estimated individually. b) Absorption efficiency of each component and the sum of all components for Pol 1. The absorption efficiency calculated from S-parameter is consistent with the one calculated from the surface loss (same for (c)).  c) Absorption efficiency of each component and the sum of all components for Pol 2.}
\label{fig:2}
\end{figure*}
In this section, the surfaces of all components of pixel are assigned with an impedance boundary in the simulation. The simulated S-parameters are shown in Fig.~\ref{fig:Spara}(b) and (c) for incident polarization 1 and 2, respectively. For the clarity of the discussion, only the modes with maximum values larger than \SI{-20}{dB} are considered and plotted. 

From the S11 of the dominant mode, the band center is \SI{-36}{dB} at \SI{272}{GHz} and \SI{-17.5}{dB} at \SI{274}{GHz} for Pol 1 and Pol 2, respectively. Compared with the design central frequency \SI{260}{GHz}, there is a \SI{12}{GHz} difference. Fig.~\ref{fig:2} (dashed curves) shows the total absorption in the pixel, as calculated from Eqn.~\ref{eqn:S11}. It has a maximum at \SI{262}{GHz} for both polarizations, which is consistent with the design. This analysis shows that actually a non-negligible part of the energy is reflected into higher modes, which are not considered in earlier studies~\cite{Roesch:2012a}.

The two dominant reflected higher modes are $\textrm{TE}_{10}$ and $\textrm{TE}_{-10}$ for Pol 1 and $\textrm{TE}_{0-1}$ and $\textrm{TE}_{01}$ for Pol 2. The maximum reflection of all higher modes together is 15.1\% and 27.2\% for Pol 1 and Pol 2, respectively. The average reflection in band from \SIrange{230}{290}{GHz} of all higher modes is 6.3\% and 13.0\% for Pol 1 and Pol 2, respectively. A large portion of the reflected higher modes is due to the reflected energy by impedance mismatch, 35.3\% at peak calculated using Eqn.~\ref{eqn:model}. We also find that the electric field distributions of these higher modes for both polarizations are consistent with the geometry of the Hilbert curve. This suggests that part of the higher modes is caused by the non-uniformity of the inductor geometry. Compared to the wavelength \SI{1.15}{mm}, the segment length of \SI{240}{\um} is quite large ($\approx \lambda /5$). Ideally the typical geometry should be much smaller than the wavelength to be treated as a uniform sheet ($< \lambda /10$)~\cite{Doyle:2008b}. However, decreasing the segment length will also decrease the inductor width, resulting in a decrease of array yield and uniformity. For example, if we increase the third-order Hilbert inductor to a fourth order, the inductor width will be around \SI{1}{\um}, and the total length of the inductor will be \SI{27}{cm}. To fabricate such long and narrow line is a challenge for the yield of fabrication. Also, this might cause distributed resonating modes coming into the band of interest.

Using the surface loss method mentioned in Sec.~\ref{sec:surface_loss}, the absorption efficiency of the individual components are calculated. Fig.~\ref{fig:2}(a) shows the definition of the four different components. The absorption efficiency calculated from S11 is consistent with the sum of the surface losses of all components (Fig.~\ref{fig:2}). The maximum difference is 1.0\% and 2.3\% for Pol 1 and Pol 2, respectively. These small differences could be caused by numerical calculation error and by small overlaps of the surfaces of connections. 

The average absorption of all components in the frequency band from \SIrange{230}{290}{GHz} is 73.1\% and 50.5\% for Pol 1 and Pol 2, respectively. The average in band absorption of the inductor is 61.7\% and 41.2\% for Pol 1 and Pol 2, respectively. There is about 10\% of the energy absorbed by other components than the inductor, for both polarizations. Compared with the calculation using Eqn.~\ref{eqn:model}, the inductor absorbs 10\% more energy for Pol 1 and 10\% less energy for Pol 2. This 20\% difference of the absorption efficiency in two polarizations was not expected from the transmission line model (Eqn.~\ref{eqn:model}). This difference of absorption efficiency in two polarizations is mainly caused by the readout line and the frame, as we will show in Sec.~\ref{sec:results 3}. The simulated absorption efficiency are summarized in Table~\ref{table}.

\subsection{FTS measurements}
\label{sec:results 2}
\begin{figure}[ht]
\begin{center}
\begin{tabular}{c}
\includegraphics[width=0.45\textwidth,trim={0cm 0cm 0cm 0cm},clip]{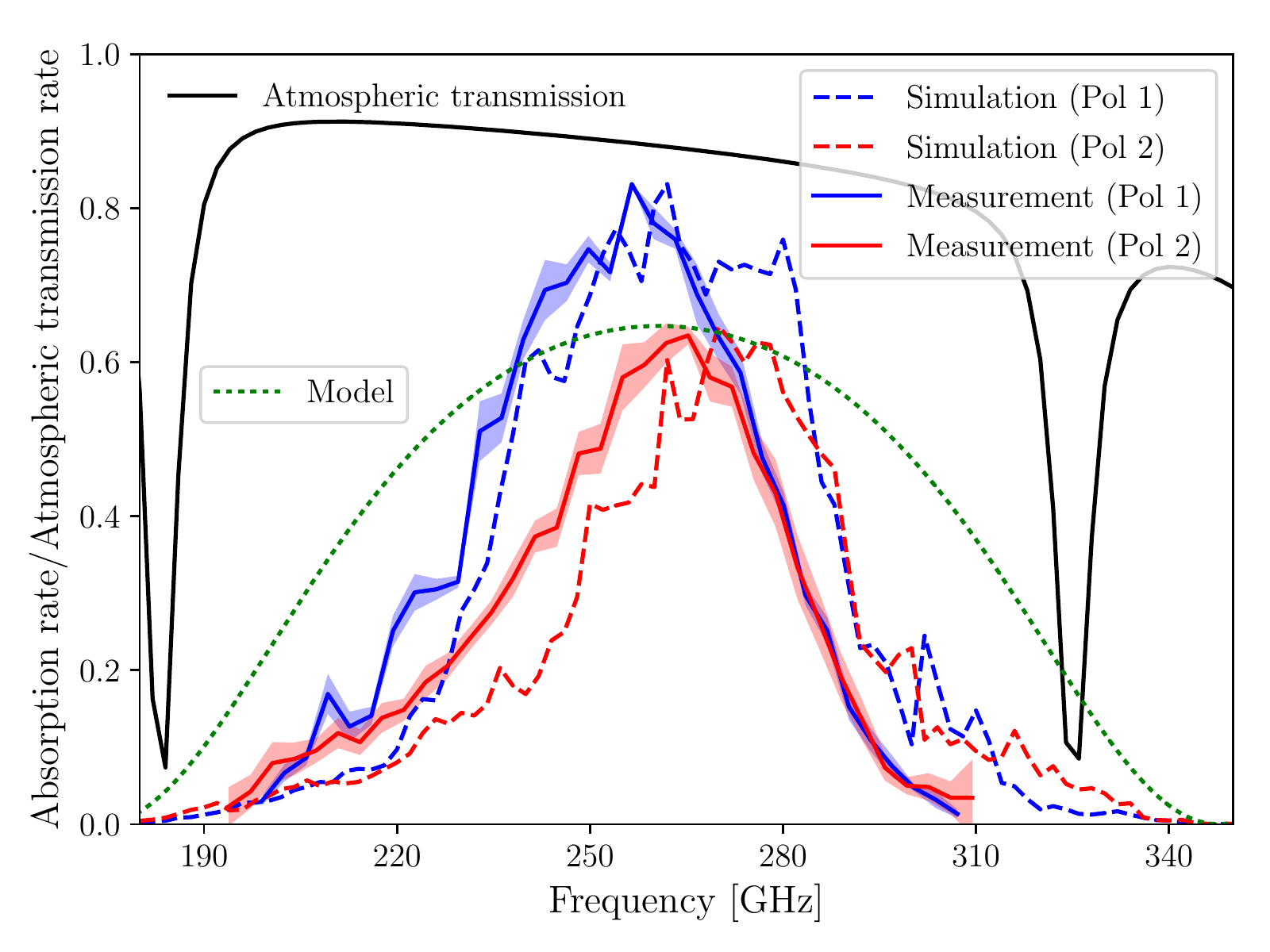}
\end{tabular}
\end{center}
\caption{Simulated absorption efficiency of the inductor for both polarizations (dashed curves) and mean (solid line) and standard deviation (shadow) of the FTS measured responses of the KIDs. The responses are normalized to the maximum of the simulation. The green dotted line is calculated from Eqn.~\ref{eqn:model}, using $Z_{\textrm{KID}}=\SI{96}{\ohm}$. The atmospheric transmission is calculated using a simplified model with precipitable water vapor of \SI{1}{mm}.}
\label{fig:FTS}
\end{figure}
\begin{table}[h]
\caption{Absorption efficiency of the NIKA2 1mm arrays}
\label{table}
\begin{center}
\begin{tabular}{|l|c|c|}
\hline
Averaged from \SIrange{230}{290}{GHz} & Pol 1 & Pol 2\\
\hline
$\textrm{AE}_{\textrm{SL}}$ of all components together & 73.1\% & 50.5\% \\
\hline
$\textrm{AE}_{\textrm{SL}}$ of other components than the inductor & 10.9\% & 9.3\% \\
\hline
$\textrm{AE}_{\textrm{SL}}$ of the inductor& 61.7\% & 41.2\% \\
\hline
Transmission line model (Eqn.~\ref{eqn:model})& \multicolumn{2}{c|}{61.5\%} \\
\hline
FTS measurement (\SIrange{220}{280}{GHz} ) & 60.0\% & 42.2\% \\
\hline
\end{tabular}
\end{center}
\end{table}
We compare the results of our simulations to measurements performed on the detector arrays fabricated for the NIKA2 project~\cite{Goupy:2016a}. To measure the optical response, two arrays are cooled down at \SI{150}{\milli\K} in the NIKA2 cryostat. A polarizer is placed in front of two arrays to separate two polarizations, Pol 1 and Pol 2. The orientation of the two polarizations with respect to the array are shown in Fig.~\ref{fig:Spara}(a). The optical passband is defined by a \SI{10.15}{cm^{-1}} (\SI{304}{GHz}) low-pass filter and a \SI{6.55}{cm^{-1}} (\SI{196}{GHz}) high-pass filter\footnotemark. The detailed optical setup is introduced in~\cite{Adam:2018a}.

\footnotetext{Both filters are from Cardiff University}

The spectral response of the arrays has been characterized using a Fourier Transform Spectrometer (FTS) based on a Martin-Puplett Interferometer~\cite{Durand:2007a} (see Appendix). The FTS uses two blackbody sources, at \SI{77}{K} and \SI{300}{K} respectively. The LEKIDs responses are measured using the NIKEL readout system~\cite{Bourrion:2016a}. The raw data is the response of each LEKID versus the position of the moving mirror. Using a Fourier transform, the position is converted to frequency. 

We measured one readout line of pixels for each array. Due to the data quality, $86$ and $112$ pixels are analyzed for Pol 1 and Pol 2, respectively. For this measurement, the maximum displacement of the mirror from the Zero Path Difference position (ZPD) is \SI{22}{mm}, resulting in a \SI{3.4}{GHz} frequency resolution. To account for the different blackbody powers emitted by the sources at different frequencies, the results are rescaled applying a $1/\nu^2$ factor, derived from the Rayleigh-Jeans limit for constant optical throughput versus frequency. The results are then calibrated using the transmission spectra of the two band-defining filters. For simplicity, the data is normalized to the peak of the simulated absorption efficiency of the inductor for each polarization. Finally the mean and standard deviation are plotted in Fig.~\ref{fig:FTS}.
\begin{figure*}[ht]
\begin{center}
\begin{tabular}{c}
\includegraphics[width=0.9\textwidth,trim={0cm 8.5cm 0cm 8.5cm},clip]{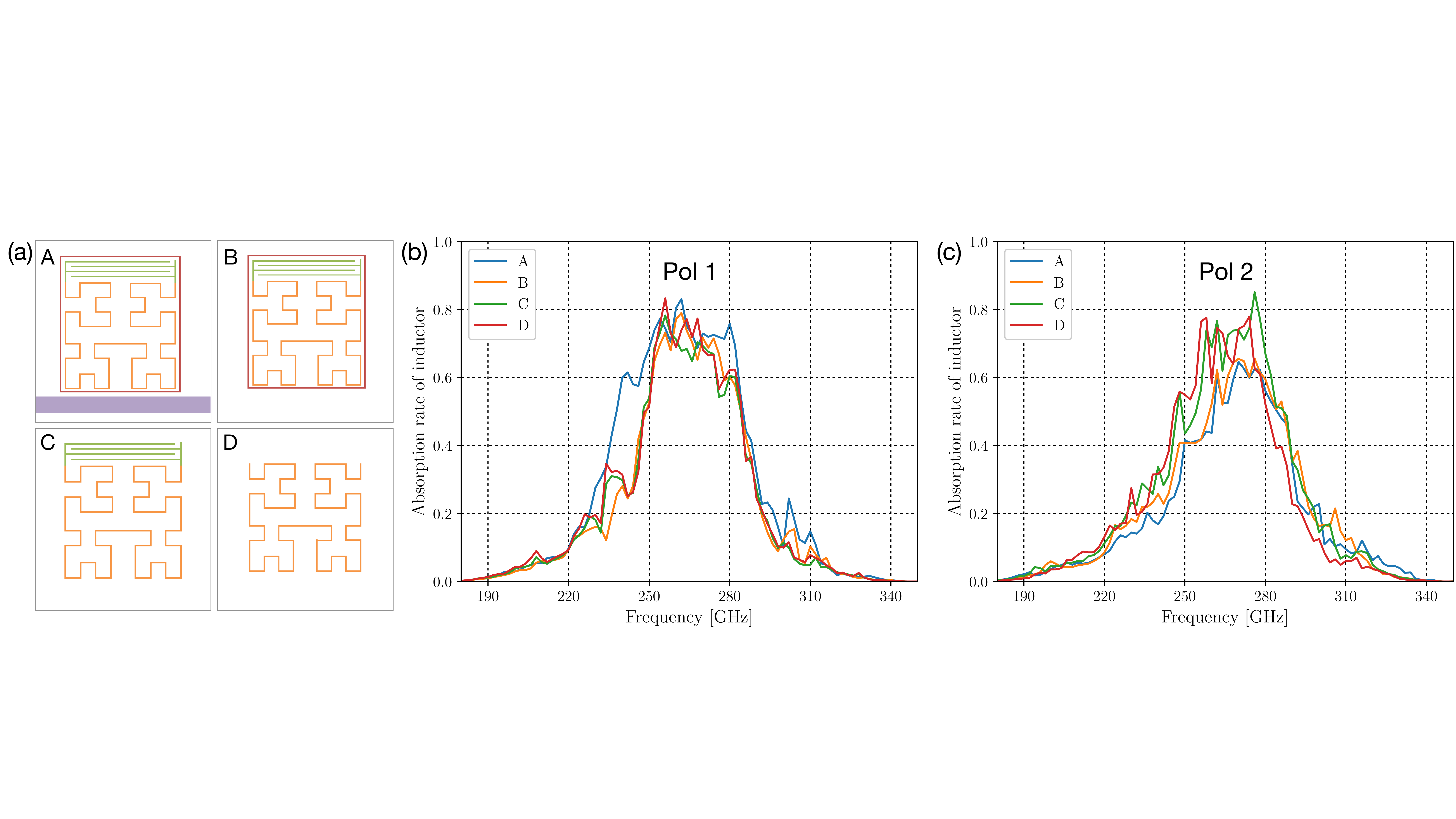}
\end{tabular}
\end{center}
\caption{a) The four simulation configurations investigated in (b) and (c). b) Absorption efficiency of the inductor, Pol 1. c) Absorption efficiency of the inductor, Pol 2.}
\label{fig:3}
\end{figure*}

To make discussion simple, we compare the simulated absorption of the inductor with the FTS measured responses of LEKIDs directly. We assume that the different lengths of the capacitor fingers in the different pixels have no influence on the absorption efficiency, since the finger length is much smaller than the wavelength and the capacitor is not sensitive to the radiation. From the FTS measurements presented in Fig.~\ref{fig:FTS}, we could also conclude that the finger lengths have no obvious influence on the optical response. The quasi-particles generated in the capacitor could change the capacitance or diffuse into the inductor. However, these changes are negligible and we therefore assume that the energy absorbed by other parts than the inductor has no effect on the response of the pixels.

From Fig.~\ref{fig:FTS}, the simulation matches well with the measurement, except for a small offset for both polarizations. This \SI{10}{GHz} offset between the measurement and simulation can be explained by the measured \SI{260}{\um} thickness of backshort for both arrays, instead of the \SI{250}{\um} used in the simulation. The substrates are commercial products with a specification of $\pm\SI{10}{\um}$. After considering this effect, we calculated the average measured response from \SIrange{220}{280}{GHz} (Table~\ref{table}) and it matches perfectly with the absorption efficiency of the inductor, discussed in Sec.~\ref{sec:results 1}. Using Eqn.~\ref{eqn:model}, the calculated absorption from the simple model gives a similar maximum with Pol 2, but overestimates the total in-band absorption efficiency. The small standard deviations of the measurement show that all pixels have a similar optical response.

\section{Further Analysis}
\subsection{Absorption efficiency by different components}
\label{sec:results 3}

To understand the results discussed in Sec.~\ref{sec:results 1}, we performed simulations for four configurations by removing individual components of the pixel while keeping everything else the same. Fig.~\ref{fig:3}(a) shows how the readout line, frame and capacitor are removed step by step in the simulation models. The average in band absorption efficiency of the inductor only decreases 10\% from A to B by removing the readout line for Pol 1. For Pol 2, the absorption efficiency increases 3\% from A to B. This indicates that the readout line focuses the electric field to the inductor, when the directions of the readout line and E-field are perpendicular to each other. When the directions are parallel, it reflects a small part of energy, like a metal mesh polarizer.

From B to C, removing the frame does not change the absorption efficiency for Pol 1, but does increases it for Pol 2 by 9\%. Also the energy reflected to the higher modes are decreased by 6\%. This suggests that the frame, used to decrease the cross talk, give rises to an increase of generation of higher modes for Pol 2. The increased absorption of the inductor in B comes from the absorbed and reflected energy of the frame in C.

From C to D, the capacitor increases the absorption of Pol 1 by 2\% and decreases that of Pol 2 by 2\%. Therefore, our compact capacitor design does not change the absorption efficiency significantly. For both C and D, the absorption efficiency for Pol 1 and Pol 2 are approximately the same, about 51\%. This is consistent with the expectation that the Hilbert curve has similar sensitivity for dual polarizations.

\subsection{Absorption efficiency with different material}
\label{sec:results 4}

\begin{figure}[ht]
\begin{center}
\begin{tabular}{c}
\includegraphics[width=0.45\textwidth,trim={0cm 0cm 0cm 0cm},clip]{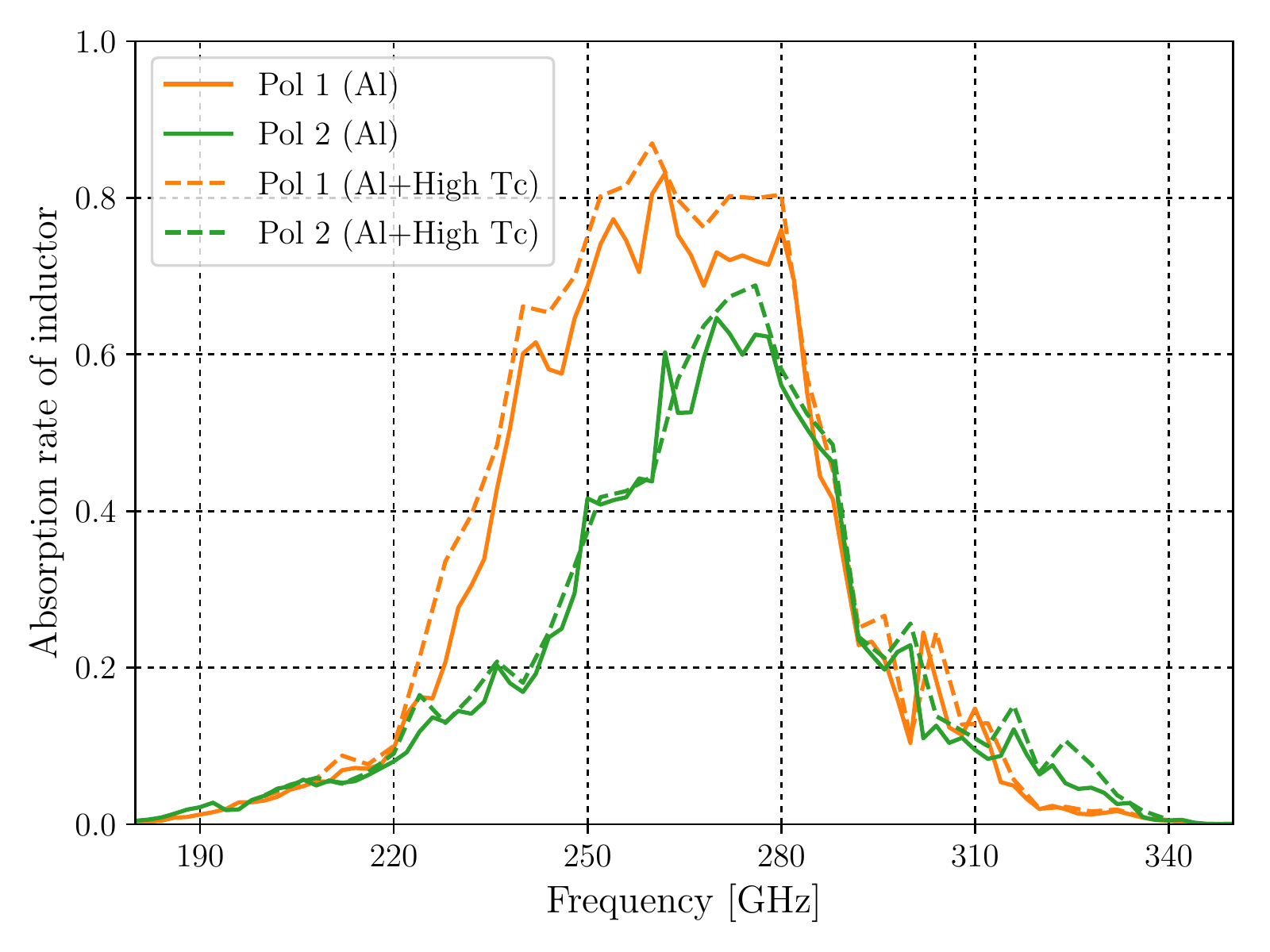}
\end{tabular}
\end{center}
\caption{Absorption efficiency of inductor with two configurations for two polarizations. All surfaces of components are assigned as impedance boundaries (Al) in one configuration. In another configuration, the surfaces of the readout line and frame are assigned as PEC.}
\label{fig:4}
\end{figure}

In Sec.~\ref{sec:results 3}, we have noticed that 10\% of the energy is absorbed by other components than the inductor. To increase the absorption efficiency of the inductor, lossless material could be used for the readout line and the frame, such as Nb, NbN, and NbTiN. These materials have a high transition temperature corresponding to high gap frequencies (\SI{73}{GHz\per K}). When the Tc of the superconductor is higher than \SI{4.5}{K}, it can be considered as a PEC in our frequency range. 

In the simulation, the surfaces of the readout line and frame are assigned as PEC while the inductor and the capacitor are still with the impedance boundary used before. It would also be possible to use a high-Tc material for the capacitor, but the connection of the inductor and capacitor could be a problem during fabrication. Using high-Tc material, the average absorption efficiency in band is increased by 7.4\% and 2.6\% for Pol 1 and Pol 2, respectively. These increases are consistent with the absorption efficiency of the readout line and frame shown in Fig.~\ref{fig:2}(b) and (c). We also simulated changing the readout line to high-Tc material only. The results have no larger than 2\% difference compared with all-Al model. This is consistent with the prediction that the sheet impedance of the wide readout line is much smaller than $Z_0$, so that it behaves as a PEC.

\section{CONCLUSIONS}

In conclusion, we have presented a detailed analysis of the optical performance of lumped-element kinetic-inductance detector arrays, using the NIKA2 1mm array as a test case. Our analysis shows that whereas a transmission line model gives a reasonable estimate of the absorption at the frequency band center, it overestimates the band-integrated optical absorption. For a Hilbert-type pixel, numerical simulations using HFSS show a significant difference in the absorption efficiency for the two polarizations, due to the anisotropic scattering off the readout line, and stress the importance of scattering to higher-order reflection modes. The detailed frequency response is validated by FTS measurements performed on the actual NIKA2 detector arrays. A detailed analysis of the absorption in different elements of the pixels show that most radiation is absorbed in the sensitive inductor part, but an improvement of 7\% can still be made by changing the frame of the pixel to a non-absorbing (high-$T_c$) material. Further improvements can be made by increasing the sheet impedance of the inductor line by decreasing the linewidth or the filling factor. By changing the geometry to a $\lambda/4$ back short, the frequency bandwidth can still be doubled. This way, a detector array that is fully optimized for continuum detection in the 1mm atmospheric window comes into reach.

\appendix[Martin-Puplett Interferometer]
\label{app}
The Martin-Puplett Interferometer (MPI) working principle is analogue to that of the Michelson interferometer: an input optical signal is split in two arms, one fixed and one of variable length. This introduces an optical path difference between the light travelling on the two arms that, once recombined, will undergo an interference, which depends on its frequency and on the length of the moving arm. This effect forms our interferogram, whose Fourier Transform makes it possible to measure the absorbed power as a function of frequency.

\begin{figure}[ht]
\begin{center}
\begin{tabular}{c}
\includegraphics[width=0.45\textwidth,trim={11cm 6cm 11cm 6cm},clip]{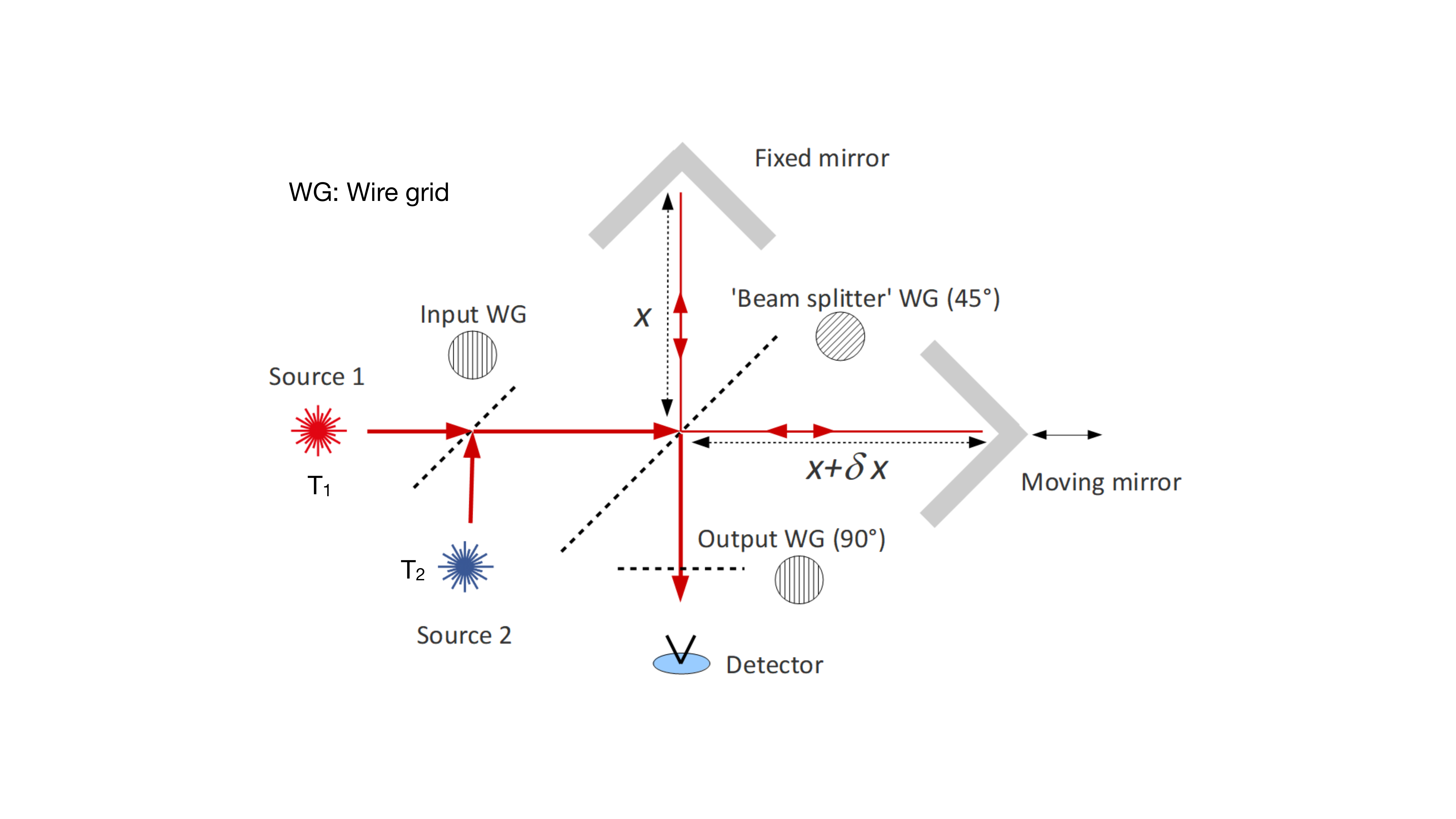}
\end{tabular}
\end{center}
\caption{The setup of our MPI. \SI{300}{K} and \SI{77}{K} blackbodies are used as hot and cold sources, respectively. Two polarizations can be measured by changing the direction of the input and output wire grids.}
\label{fig:MPI}
\end{figure}

In practice, an MPI is based on wire grid polarizers, shown in Fig.~\ref{fig:MPI}. At the input, a first polarizer is mounted at 45$^\circ$ inclination between two blackbody sources at different temperatures, $T_1$ and $T_2$. This signal then reaches the central polarizer, where it is split in the two different arms. The mirrors at the end of the arms are roof top mirrors, introducing a 90$^\circ$ shift in the polarization of the signal. Thus, when reaching once more the central polarizer, the polarization that was reflected at the first interaction is now transmitted, and vice versa. The output signal is the sum of this two polarizations. A last wire grid on the output of the MPI makes it possible to detect the interferogram. One can show that, using the appropriate inclination of the various wire grids, the expected signal after the output wire grid is given by
\begin{equation}
\begin{split}
P_\nu (\delta x) = & \frac{1}{2}\left( B_\nu(T_1) + B_\nu(T_2)\right) \\
& + \frac{1}{2} \left(B_\nu(T_1) - B_\nu(T_2)\right)\cos(2\pi\nu\cdot 2\delta x/c).
\end{split}
\end{equation}
$P_\nu(\delta x)$ is the amount of output power at frequency $\nu$ when the moving mirror is at a distance $\delta x$ from the ZPD position, and $B_\nu(T)$ is the brilliance of a blackbody at temperature $T$. The first term can be removed by making the first wire grid rotate around its axis and applying a lock-in technique to get the LEKID signal. The observed interferogram, $I(\delta x)$, will be given by 

\begin{equation}
I(\delta x) = \int_{\nu_{min}}^{\nu_{max}}P_\nu(\delta x)A_\nu d\nu
\end{equation}

where $A_\nu$ is the absorption of the considered LEKID. The Fourier Transform is applied to the interferogram to finally get $A_\nu$.

The maximum displacement of the moving arm, $\Delta x$, determines the resolution of the measured spectrum, whereas the smallest step by which the moving mirror can be controlled impacts the maximum detectable frequency. In our system we usually span about \SI{20}{mm} around the ZPD position, so the resolution is of $\sim \SI{3}{GHz}$. The resolution of the stepper motor is $\sim \SI{20}{\um}$, meaning that the spectrum can be measured in principle up to a few \si{THz}.

\section*{ACKNOWLEDGMENT}

We thank Alvaro Gonzalez, Tom Nitta, Wenlei Shan and Reinier M.J. Janssen for useful discussion and Yutaro Sekimoto for encouragement. We thank Florence Levy-Bertrand for contributing to the maintenance of the FTS, and Alain Benoit for the construction of the FTS and the software for these measurements. The HFSS simulations were performed at National Astronomical Observatory of Japan.


\bibliographystyle{IEEEtran}
\bibliography{Shu_lib.bib}
\begin{biography}[{\includegraphics[width=1in,height=1.25in,clip,keepaspectratio]{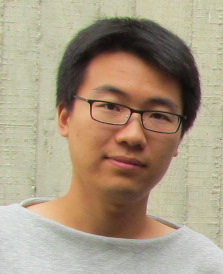}}]{Shibo Shu}
received the B.S. degree in electronic science and technology from the University of Electronic Science and Technology of China, China, in 2013, the M.Sc. degree in astronomy from the University of Tokyo, Japan, in 2015, and is currently a Ph.D. student in physics at the Institut de RadioAstronomie Millim\'{e}trique and the Universit\'{e} Grenoble Alpes, France.

From 2013 to 2016, he received the Japanese Government Scholarship and worked on kinetic inductance detectors for the LiteBIRD satellite and the Nobeyama 45-m telescope at the National Astronomical Observatory of Japan. From 2016, he started to work in the NIKA2 group for the Ph.D. degree.
\end{biography}

\begin{biography}[{\includegraphics[width=1in,height=1.25in,clip,keepaspectratio]{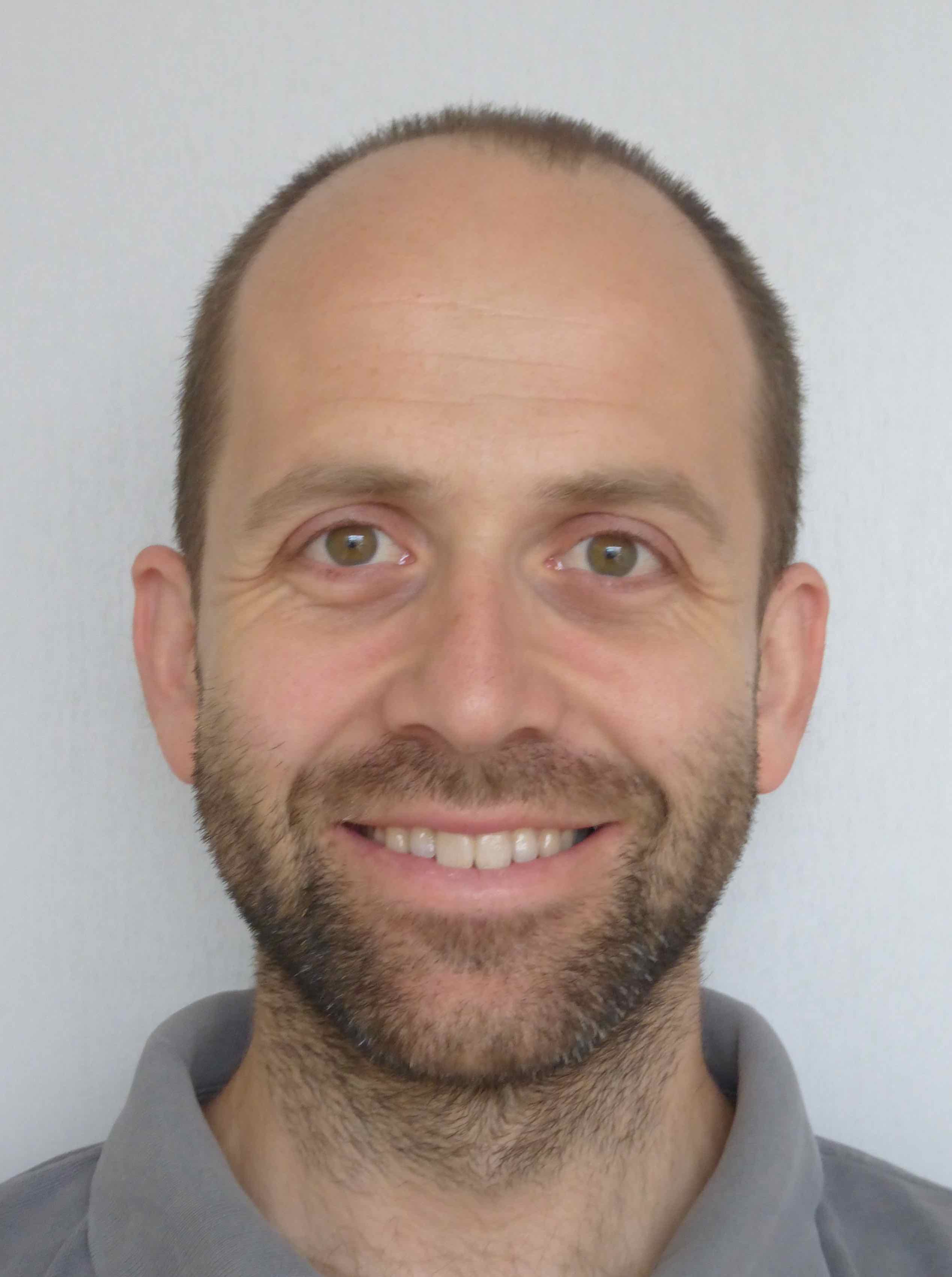}}]{Martino Calvo}
received his Graduate and PhD degree at the University of Rome Sapienza in 2005 and 2009 respectively. His PhD thesis was devoted to the developement of the first Lumped Element KIDs dedicated to millimeter wave detection. He continued his work on KIDs during a 2 years Post-Doc grant at the same university, before moving to France where he joined the Neel Institute of CNRS Grenoble in 2011. Since then, he has been involved in the development and realisation of various instruments for mm-wave astronomy, in particular NIKA2, the first KID-based camera to observe the sky. Since 2016 he is Research Engineer at Institut Neel.
\end{biography}

\begin{biographynophoto}{Johannes Goupy}
, photograph and biography not available at time of publication.
\end{biographynophoto}

\begin{biographynophoto}{Andrea Catalano}
received Ph.D. degree from the Observatoire de Paris/Meudon in December 2008. His dissertation was entitled "Development of digital models of the High Frequency Instrument (HFI) of Planck needed for its operation". Today He is currently researcher at CNRS working at the laboratoire de Physique Subatomique et Cosmologie (LPSC) in Grenoble. 
\end{biographynophoto}

\begin{biographynophoto}{Aurelien Bideaud}
, photograph and biography not available at time of publication.
\end{biographynophoto}

\begin{biography}[{\includegraphics[width=1in,height=1.25in,clip,keepaspectratio]{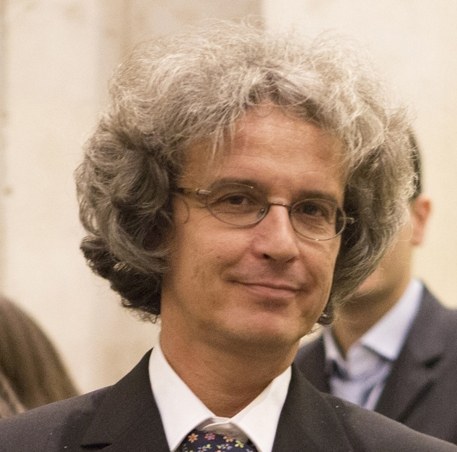}}]{Alessandro Monfardini}
(born in Brescia, Italy, in 1971) is a physicist at CNRS and Research Director of the Néel Astrophysics Instrumentation group. Since his master degree in 1996 at the University of Trieste, he has developed systems, instruments, detectors and software for a number of applications including astrophysics, particle and nuclear physics, and space studies. He has worked in Italy, Japan, UK and France.
\end{biography}

\begin{biographynophoto}{Samuel Leclercq}
received a physicist PhD degree at the University of Grenoble, France, in 2004. His thesis topic being the development of a bolometers based camera for radioastronomy at millimeter wavelengths. His Post-Doc at the university of Florida, USA, between 2004 and 2006, was dedicated to data analysis, simulations ans instrument operations as a member of the Cold Dark Matter Search (CDMS) collaboration. Since 2006 he works at IRAM as optics designer, project scientist and coordinator mostly involved in the development of the NIKA2 instrument, installed at the IRAM 30m telescope in Spain. 
\end{biographynophoto}

\begin{biographynophoto}{Eduard F. C. Driessen}
received his MSc degree in Applied Physics from Delft University of Technology, Delft, The Netherlands, in 2005, and his PhD degree from Leiden University, Leiden, The Netherlands, in 2009. He was a Post-doctoral fellow at Delft University of Technology from 2009-2012 and at the Commissariat à l'Energie Atomique et aux Energies Alternatives in Grenoble, France, 2012-2014. Since 2015, he works at the Institut de RadioAstronomie Millim\'{e}trique in Grenoble, France, where he leads the Superconducting Devices Group, and is in charge of the development of heterodyne and continuum superconducting detectors for millimeter-wave radioastronomy.
\end{biographynophoto}
\end{document}